\documentclass[
 aps,
 rsi,
 superscriptaddress, 
 twocolumn
]{revtex4-2}

\usepackage{graphicx}% Include figure files
\usepackage{dcolumn}% Align table columns on decimal point
\usepackage{bm}% bold math
\usepackage[utf8]{inputenc}
\usepackage[T1]{fontenc}
\usepackage{mathptmx}
\usepackage{etoolbox}
\usepackage{siunitx}

\makeatother

\begin{document}

\preprint{AIP/123-QED}

\title[Sample title]{A Time-Resolved Imaging System for the Diagnosis of X-ray Self-Emission in High Energy Density Physics Experiments}
% Force line breaks with \\
\author{Jack W. D. Halliday}
	\email{jack.halliday12@imperial.ac.uk}
	\address{Blackett Laboratory, Imperial College, London SW7 2BW, United Kingdom}%Lines break automatically or can be forced with \\
 	
\author{Simon N. Bland}
	\address{Blackett Laboratory, Imperial College, London SW7 2BW, United Kingdom}
\author{Jack. D. Hare}
	\address{Department of Nuclear Science and Engineering, Massachusetts Institute of Technology, 77 Massachusetts Avenue, Cambridge, MA 02139}
\author{Susan Parker}
	\address{Blackett Laboratory, Imperial College, London SW7 2BW, United Kingdom}

\author{Lee G. Suttle}
	\address{Blackett Laboratory, Imperial College, London SW7 2BW, United Kingdom}
\author{Danny R. Russell}
	\address{Blackett Laboratory, Imperial College, London SW7 2BW, United Kingdom}
\author{Sergey V. Lebedev}
	\address{Blackett Laboratory, Imperial College, London SW7 2BW, United Kingdom}

\date{\today}% It is always \today, today,

\begin{abstract}
A diagnostic capable of recording spatially and temporally resolved X-ray self emission data was developed to characterise experiments on the MAGPIE pulsed-power generator. The diagnostic used two separate imaging systems: A pinhole imaging system with two dimensional spatial resolution and a slit imaging system with one dimensional spatial resolution. The two dimensional imaging system imaged light onto image plate. The one dimensional imaging system imaged light onto the same piece of image plate and a linear array of silicon photodiodes. This design allowed the cross-comparison of different images, allowing a picture of the spatial and temporal distribution of X-ray self emission to be established. The design was tested in a series of pulsed-power driven magnetic-reconnection experiments.       
\end{abstract}

\maketitle

\section{\label{Intro}Introduction}

The X-ray emission from high energy density (HED) plasmas is typically both transient and exhibits significant spatial variation. Traditionally, X-ray imaging data is captured either using gated X-ray detectors (for example micro-channel-plates \cite{Bland2004}) or time-integrated detectors (such as image plate \cite{Meadowcroft2008}). Time-gated detectors provide unambiguous information about the time interval in which X-ray emission occurs however they are expensive, bulky, introduce significant experimental complexity. Time-integrated detectors are compelling in the sense they are simple and inexpensive, however they provide no information about the time interval in which X-ray emission occurs and some dynamical aspects of experiments are also obscured by motion blurring. 

The information gained from time-integrated X-ray detectors is often augmented with (time-resolved) detectors which convert X-ray flux into an electrical signal. Examples of such detectors include PCDs \cite{Spielman1997}, vacuum X-ray diodes (XRDs)\cite{Dewald2004}, and silicon photodiodes. These are however typically fielded as spatially-integrated diagnostics.            

In this paper, we describe a diagnostic which bridges the gap between the time-integrated and time-resolved X-ray diagnostics discussed above. This novel design made use of two separate X-ray imaging systems: A pinhole which imaged light onto image plate (time integrating), and a slit which imaged light onto both image plate and a linear array of silicon diodes (time-resolved). This meant that two-dimensional time-integrated, and one-dimensional time-resolved images were captured simultaneously. Features in the two-dimensional and one-dimensional images could be compared. 

An approach similar to the one described here has previously been used to diagnose gas puff implosions on pulsed-power machines including Saturn and Double-EAGLE \cite{Parasad2004, Coleman2010} however time integrating X-ray detectors were not fielded alongside time-resolved ones.

There is an obvious analogy between the diagnostic we describe and an X-ray streak camera (which could also be set up to provide spatial resolution in one dimension). That said, much like MCP cameras, X-ray streaks are bulky, complex, and expensive diagnostics to field. They also have notable issues regarding both dynamic range and linearity of response. By contrast, the arrangement of silicon photodiodes we present is cheap, compact, and has a simpler response characteristic. That said, the diodes provide a coarser level of spatial resolution than an X-ray streak camera so should be seen as a complimentary technology which is best suited to harsh diagnostic environments, or situations in which a compact setup is required.             

The design we describe could be applied to the diagnosis of a wide variety of HED plasmas, although it is best suited to systems which are either extended in one dimension or have a quasi one dimensional geometry. Examples of experiments which are extended in one dimension include gas-puff or wire-array Z-Pinch implosions. Examples of experiments with a quasi one dimensional geometry include the work on photoionisation fronts which currently form part of the work done by the Z Astrophysical Plasma Properties (ZAPP) collaboration \cite{LeFevre2021}.      

The design we present was used to diagnose X-ray self emission in reconnection experiments performed on the MAGPIE generator at Imperial College \cite{Mitchell1996}. A diagnostic based on the design described here is also being developed to diagnose plasma dynamics in magnetic reconnection experiments performed on the Z-Machine at Sandia National Laboratories \cite{Savage2011}. 

\section{Optical Design of the instrument}
\begin{figure*}
	\centering
	\includegraphics[width=\textwidth]{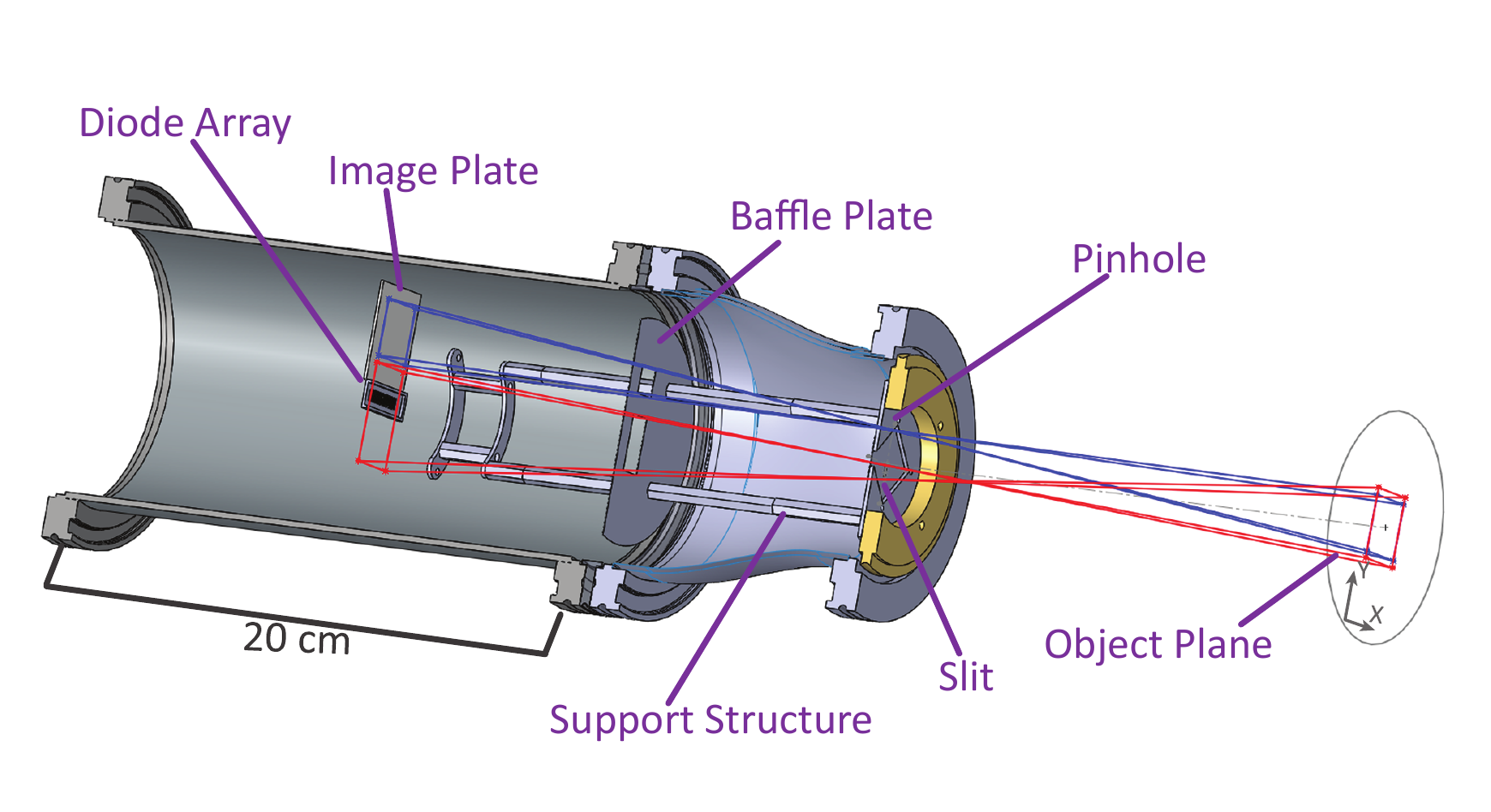}
	\caption{
		A cross section of the diagnostic which was use to image X-ray emission onto a diode array and an image plate. Ray traces for the pinhole and slit imaging systems are shown in blue and red respectively. The spatial extent of the blue and red rectangles in the object and image planes are defined by the projection of the pinhole/slit and the baffles. 
	}
	\label{fig:diode_CAD}
\end{figure*}

Figure \ref{fig:diode_CAD} shows a CAD model of the X-ray imaging system our diagnostic employed. The system used two separate apertures to image radiation from the experimental volume. The first of these was a pinhole which cast light (blue rays in figure \ref{fig:diode_CAD}) onto a piece of image plate positioned on the same plane as the diode array. The second was a slit imaging system which cast light (red rays in figure \ref{fig:diode_CAD}) onto both the image plate and the diode array. The slit imaging system provided one dimensional spatial resolution whilst the pinhole provided two dimensional resolution. 

\begin{figure}
	\centering
	\includegraphics[width=\columnwidth]{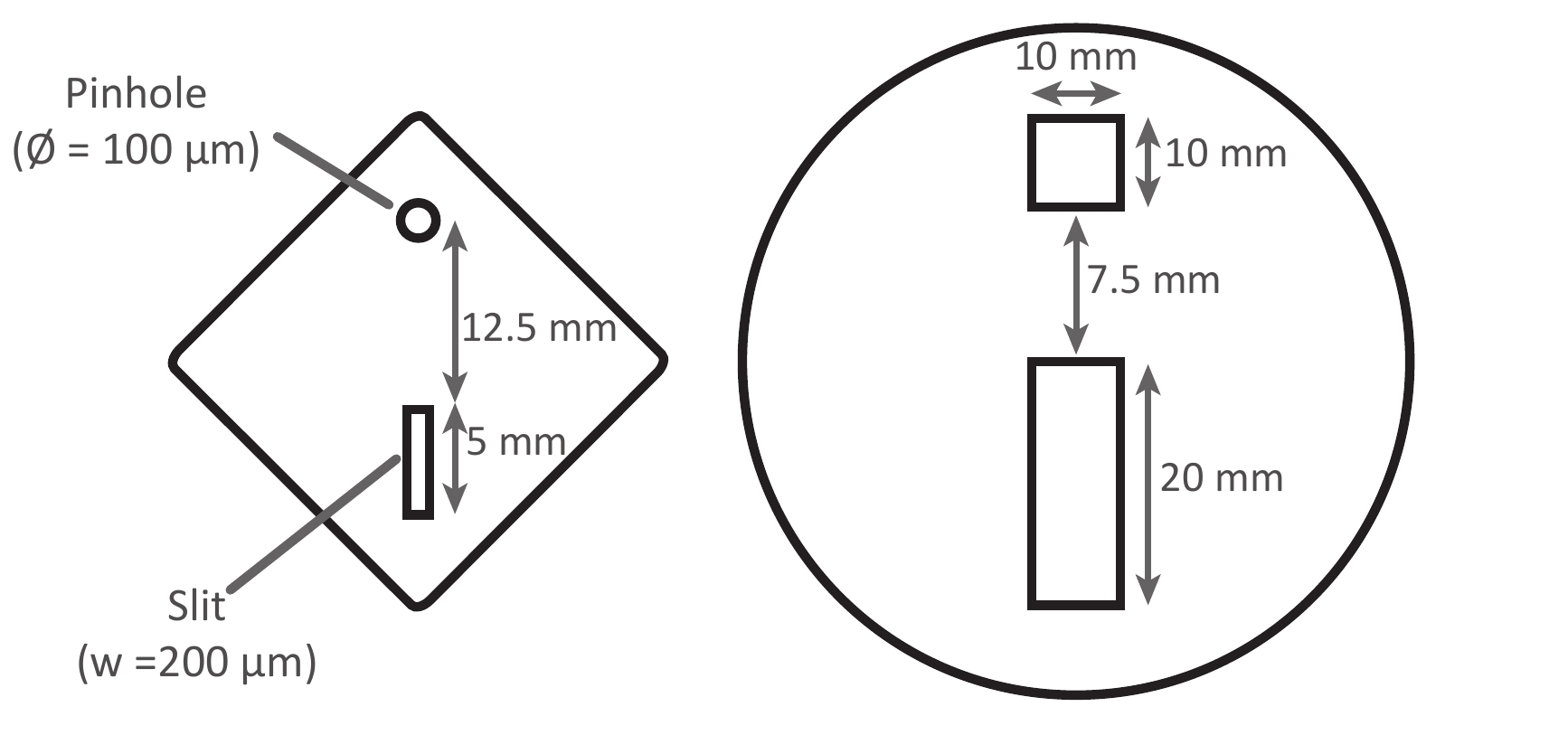}
	\caption{
		\emph{Left:} A diagram (not to scale) of the plate containing the X-ray apertures used to image the reconnection plane. \emph{Right:} A diagram of the baffle plate, which was designed to prevent cross talk between the two imaging systems used in this setup. Both plates were laser cut from \SI{200}{\um} thick stainless steel. 
	}
	\label{fig:diode_apatureDiagram}
\end{figure}

Diagrams showing the geometry of the X-ray apertures, and the baffling which was employed to prevent cross talk between the two imaging systems are shown in figure \ref{fig:diode_apatureDiagram}. Both of these components were laser-cut from \SI{200}{\micro\meter} thick stainless steel, which was sufficiently opaque to block the relatively soft X-ray emission encountered in experiments performed on MAGPIE (nominal color temperature for emission is typically $\sim \SI{100}{\electronvolt}$). Inspection of laser cut components under a microscope revealed that features were accurate to $\sim \SI{20}{\micro\meter}$.    

\begin{figure}
	\centering
	\includegraphics[width=\columnwidth]{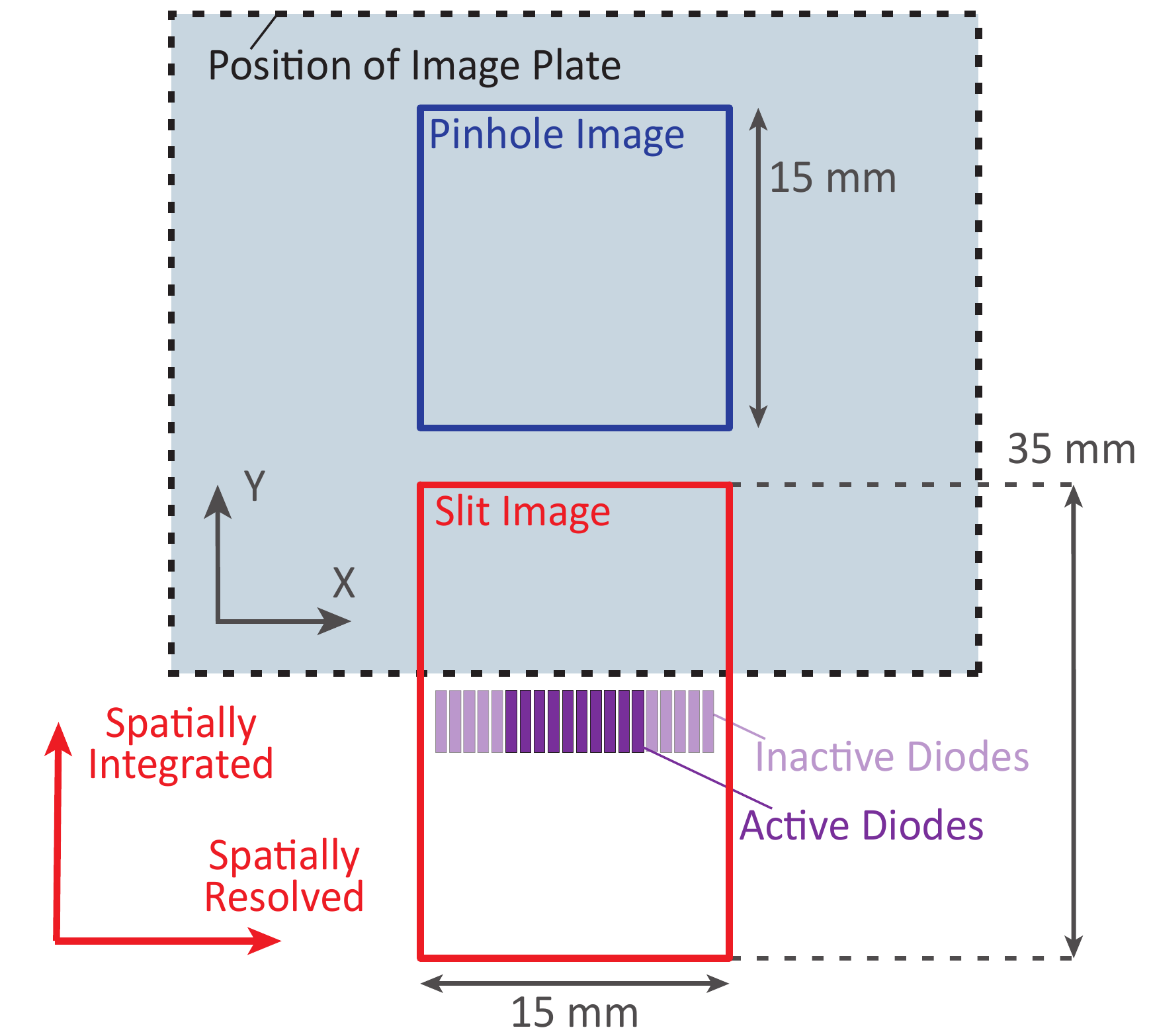}
	\caption{
		A diagram showing images formed on the detector plane by the setup used for experiments with a linear photodiode array. The diagram shows the setup formed two images, a slit image and a pinhole image. The diode array consisted of 20 photoconducting elements, the central 10 of which were used in experiments -- with the rest shorted out.  
	}
	\label{fig:diode_image_plane}
\end{figure}

The projection of the ray traces onto the image plane is shown in figure \ref{fig:diode_image_plane}. This diagram also shows the relative positions of the image plate and the diode array.  The diode array was oriented so that it was spatially resolving in the same direction as the slit.  The system was set up so that, for light cast from the object plane, the X-axis of the pinhole image could be directly related to the spatially resolved axis of the slit image. This enabled features seen in the signal from the diode array to be related to features seen in diagnostic images obtained in separate experiments. 

\section{Details of X-ray detection technology}
Time-integrated data was captured using Fuji-film image plate (BAS-TR). The image plates were scanned using a Fuji-Film FLA-5000 scanner.

\begin{figure}
	\centering
	\includegraphics[width=\columnwidth]{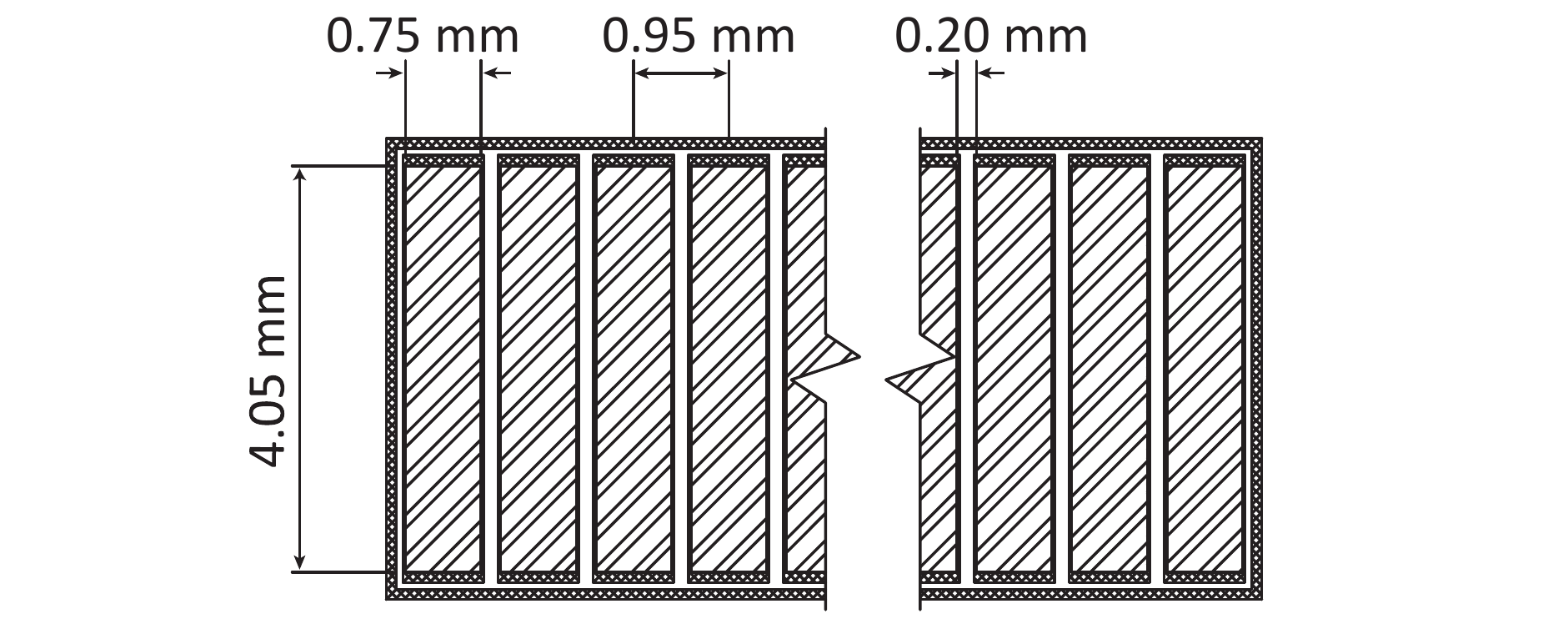}
	\caption{
		A diagram showing the geometry of the diode array. Photoconducting elements are indicated by diagonal shaded features in the diagram. The array incorporated a total of 20 photoconducting elements. 
	}
	\label{fig:diode_geometrey} 
\end{figure}
The diode-array was an Opto Diode product with the model number AXUV20ELG. This consists of 20 photoconducting elements arranged linearly, as shown in figure \ref{fig:diode_geometrey}.  The size of each element is $0.75\times\SI{4.05}{\mm\squared}$ and that the centre-center separation between elements is \SI{0.95}{\mm}. This detector is silicon based and the thickness of the active layer is \SI{50}{\micro\meter}. The spatial resolution was dominated by the geometric factors rather than diffractive effects. The width of the slit was chosen so the resolution was $\sim \SI{1}{\milli\meter}$ which is comparable to the separation between the photoconducting elements.

This diode array has a common cathode, and an individual anode for each photoconducting element. The cathode was biased to \SI{9}{\volt} and the current through each element was relayed to a channel on an oscilloscope (HP Agilent 16500B with five 16532A digitizing oscilloscope cards) via a coaxial line terminated with a \SI{50}{\ohm} resistor. A passive low pass filter ($RC = \SI{10}{\kilo\ohm} \times \SI{220}{\nano\farad}$) was used on the (common) cathode in order to reduce the effects of electrical noise from the pulsed power generator and to maintain the bias voltage on the timescale of an experiment. No filtration was used between the anode side of the diodes and the channels of the oscilloscope. 

The rise time of the detectors were empirically determined to be $\sim \SI{1}{\nano\second}$ by measuring their $V(t)$ response to radiation from a \SI{200}{\pico\second} pulsed laser (AXUV diodes have a low-level response in the visible). The discharge time for the capacitor is around $\SI{50}{\ohm} \times \SI{220}{\nano\farad} = \SI{11}{\micro\second}$ so the bias voltage was not significantly depleted on the timescale the an experiment (less than \SI{0.5}{\micro\second}). 

The hydrodynamic timescale for typical experiments on MAGPIE is $\sim \SI{10}{\nano\second}$, so the temporal response of the detector is sufficient to resolve dynamics in our experiments. This temporal response is set (in part) by the bias voltage applied to the diodes and the bias voltage we used was selected to provide a sufficiently short temporal response without causing the diodes to breakdown. 

We note that an alternative biasing scheme would have been to hold the cathode at ground, and have the bias voltage relayed to each individual photoconducting element, similar to the design reported in \cite{Spielman1997}. This alternative biasing scheme would reduce the possibility of cross-talk between the diodes in the array. 

As previously stated AXUV20ELG detectors use a \SI{50}{\micro\meter} thick active layer of silicon. The detector exhibits a good level of quantum efficiency from $\sim \SI{100}{\electronvolt}$, up to $\sim \SI{10}{\kilo\electronvolt}$. For X-rays harder than this upper limit, the penetration of the (relatively thin) active layer limits the effective quantum efficiency.

\section{Application pulsed-power driven magnetic reconnection experiments}
To validate the utility of the design described above, we fielded a prototype diagnostic in a series of pulsed-power driven magnetic reconnection experiments. Results from this magnetic reconnection platform have been previously reported in a number of publications \cite{Suttle2016, Hare2017d, Hare2017b, Suttle2018, Hare2018}. The setup was chosen here as the conditions of the plasmas produced are well characterised and the experimental morphology is extended in one dimension (rendering it suitable for diagnosis with the diode array).   
\begin{figure}
	\includegraphics[width=\columnwidth]{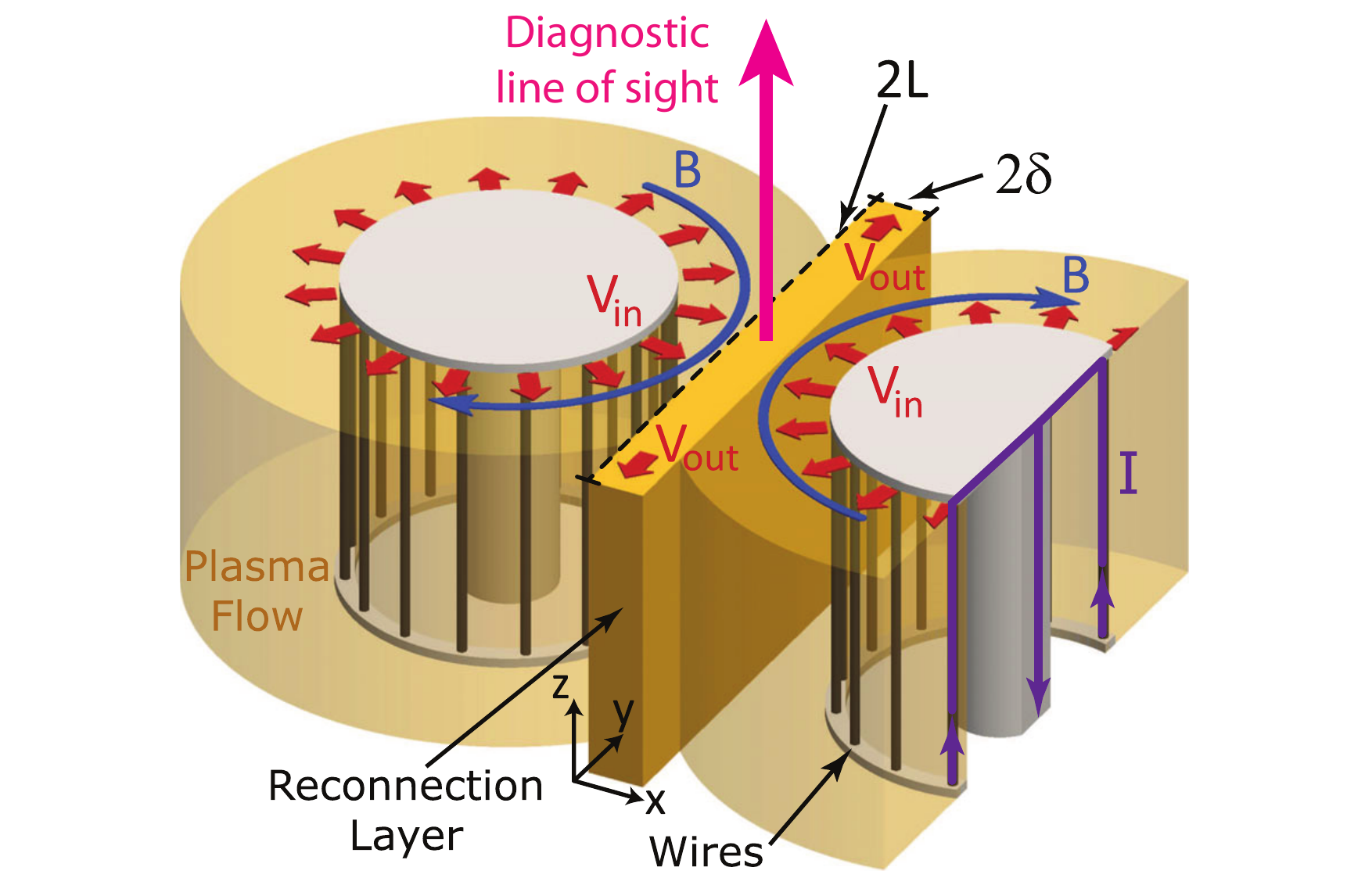}
	\caption{\label{fig:reconnection_platform} Illustration of a pulsed power driven magnetic reconnection experiment. The direction of current flow is drawn in purple; the orientation of magnetic fields are drawn in blue; and the direction of velocities are drawn in red. Reprinted  with permission from J. D. Hare et. al, Physics Review Letters, 118, 085001 (2017). Copyright 2017 by the American Physical Society.}
\end{figure}

The experimental setup is shown in figure \ref{fig:reconnection_platform}. It consists of two inverse (or exploding) wire arrays \cite{Harvey-Thompson2014} which are driven in parallel by an intense current pulse. Each array acts as a radially diverging source of plasma carrying an azimuthal magnetic field. These plasma flows are continually driven for the duration of the generator's current pulse. For experiments on MAGPIE (\SI{1.4}{\mega\ampere}, \SI{500}{\nano\second} current pulse), the typical conditions in the ablated plasma are $n_e \sim 10^{17} \; \si{\per\centi\meter\cubed}$; $T_e \sim \SI{15}{\electronvolt}$; $B_y \sim \SI{5}{\tesla}$.    

As shown in the figure, the two plasma flows from the two arrays collide in a central interaction region, forming a reconnection layer. Where the two flows meet they carry oppositely directed magnetic fields. The magnetic fields reconnect, and magnetic energy is dissipated heating and accelerating the plasma flows. Within the central reconnection layer, the temperature and electron density of the plasma increase by around an order of magnitude. The spatial scale of the reconnection layer is typically $\SI{1}{\milli\meter} \times \SI{20}{\milli\meter}$.
\begin{figure}
	\centering
	\includegraphics[width=\columnwidth]{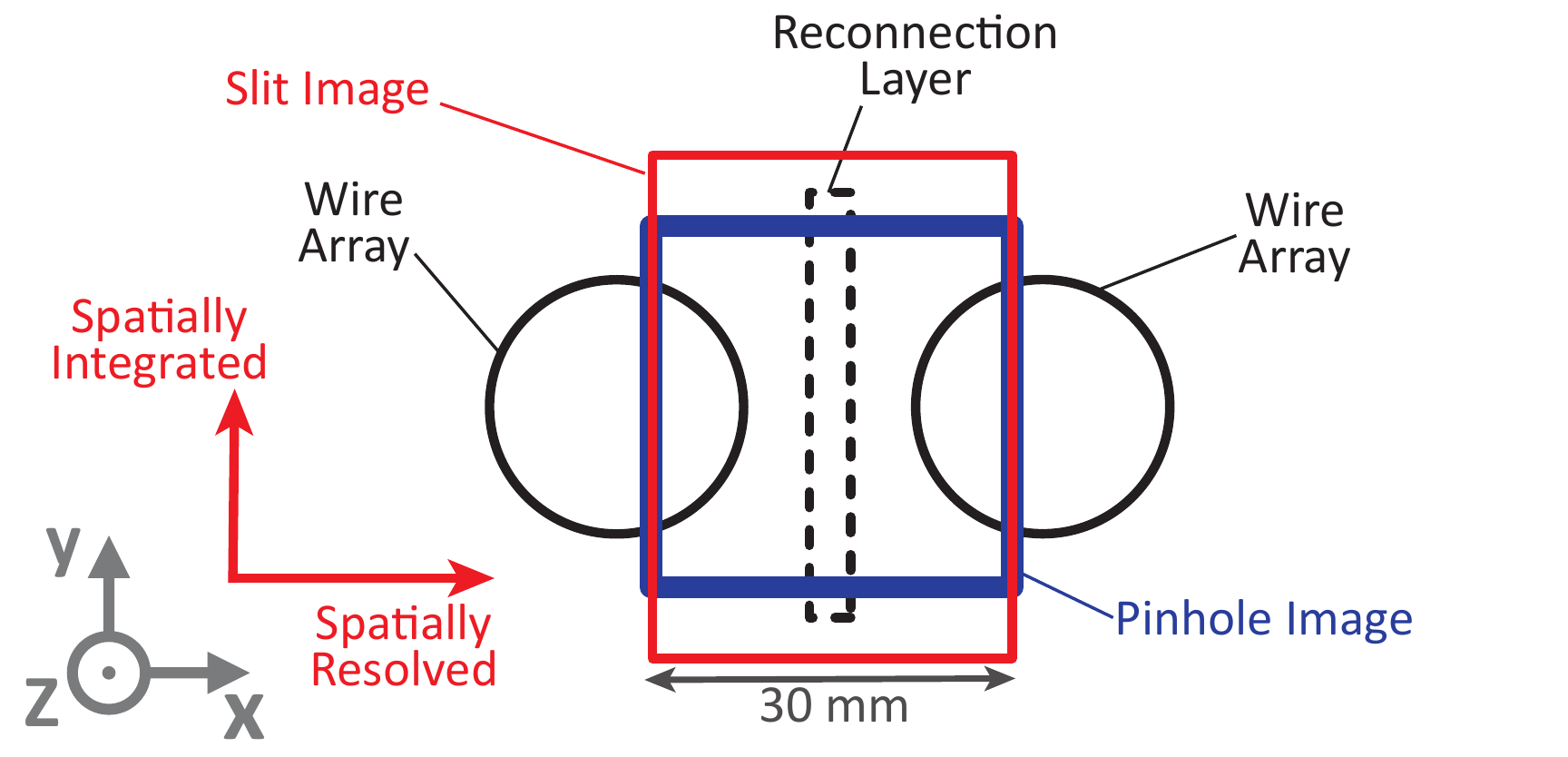}
	\caption{
		Diagram showing the image plane with projected positions of features in the reconnection plane of the experiment.
	}
	\label{fig:diode_object_plane}
\end{figure}

Figure \ref{fig:diode_object_plane} shows the regions of the experimental volume which were imaged to the diode array / image plate for the setup used in diagnostic tests. The figure shows that the slit was oriented to resolve structure across the reconnection layer (in the X-direction of the experiment). The magnification was $M=1$, so the width of the layer on the image plane was expected to be $\sim \SI{1}{\milli\meter}$. 

\begin{figure}
	\centering
	\includegraphics[width=\columnwidth]{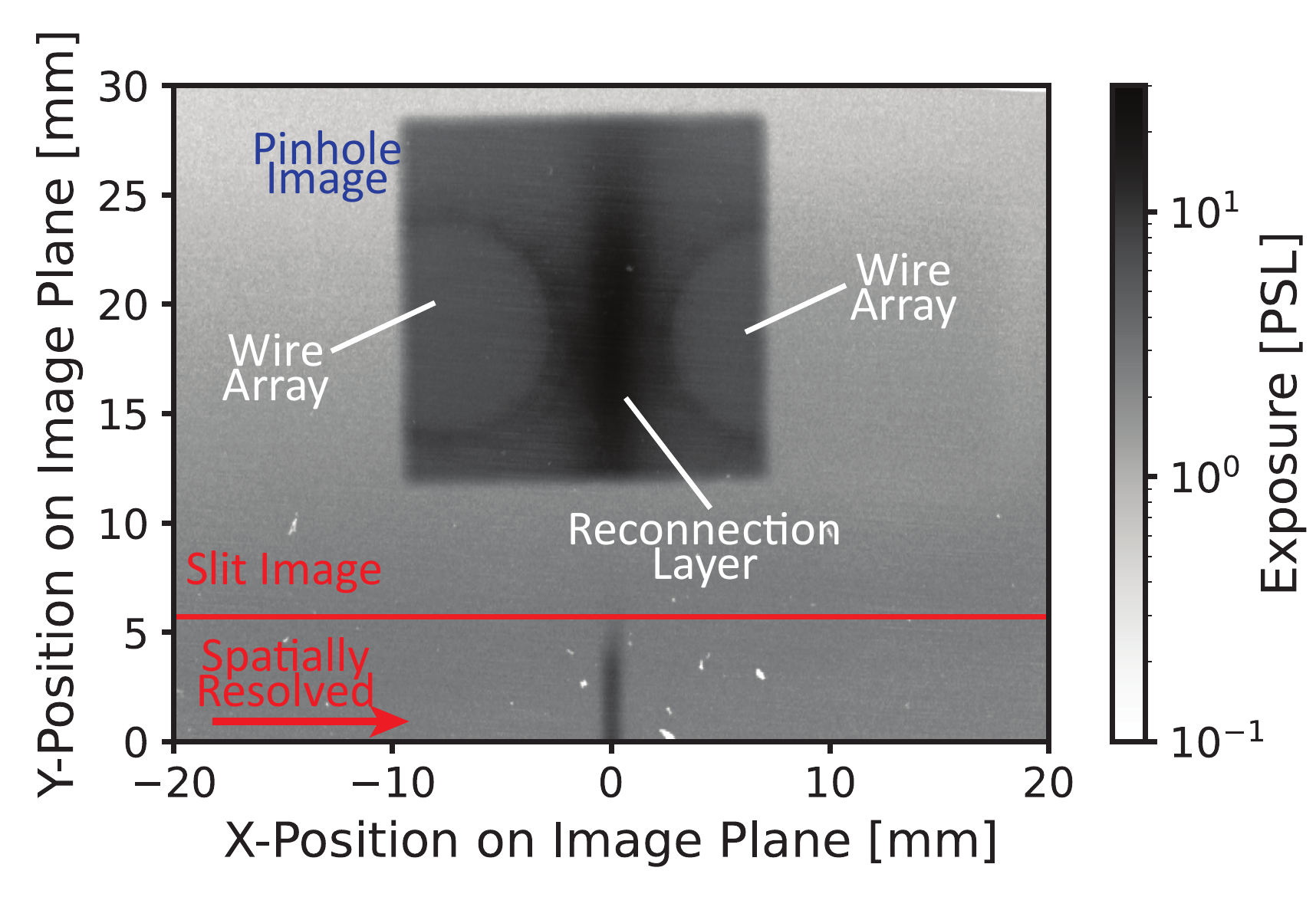}
	\caption{
		Image plate data from an experiment in which the linear photodiode array was fielded. The extent of the slit image (which was predicted through raytracing) is shown as a red horizontal line. The image is plotted in units of photostimulated luminescence (PSL), which is a measure of energy deposited in the image plate.   
	}
	\label{fig:diode_image_plate}
\end{figure}

Figure \ref{fig:diode_image_plate} shows some example image plate data obtained using this experimental setup. For diagnostic tests, the pinhole was filtered with \SI{40}{\micro\meter} polypropylene and the slit was filtered with \SI{6.5}{\micro\meter} aluminium. The polypropylene filter transmits radiation from the both the plasma inflows and the reconnection layer whereas the aluminium transmits only the hardest radiation produced in the experiment. This means that the slit image only shows signal from the hottest plasma in the experiment, which is lies within the reconnection layer. 

\begin{figure}
	\centering
	\includegraphics[width=\columnwidth]{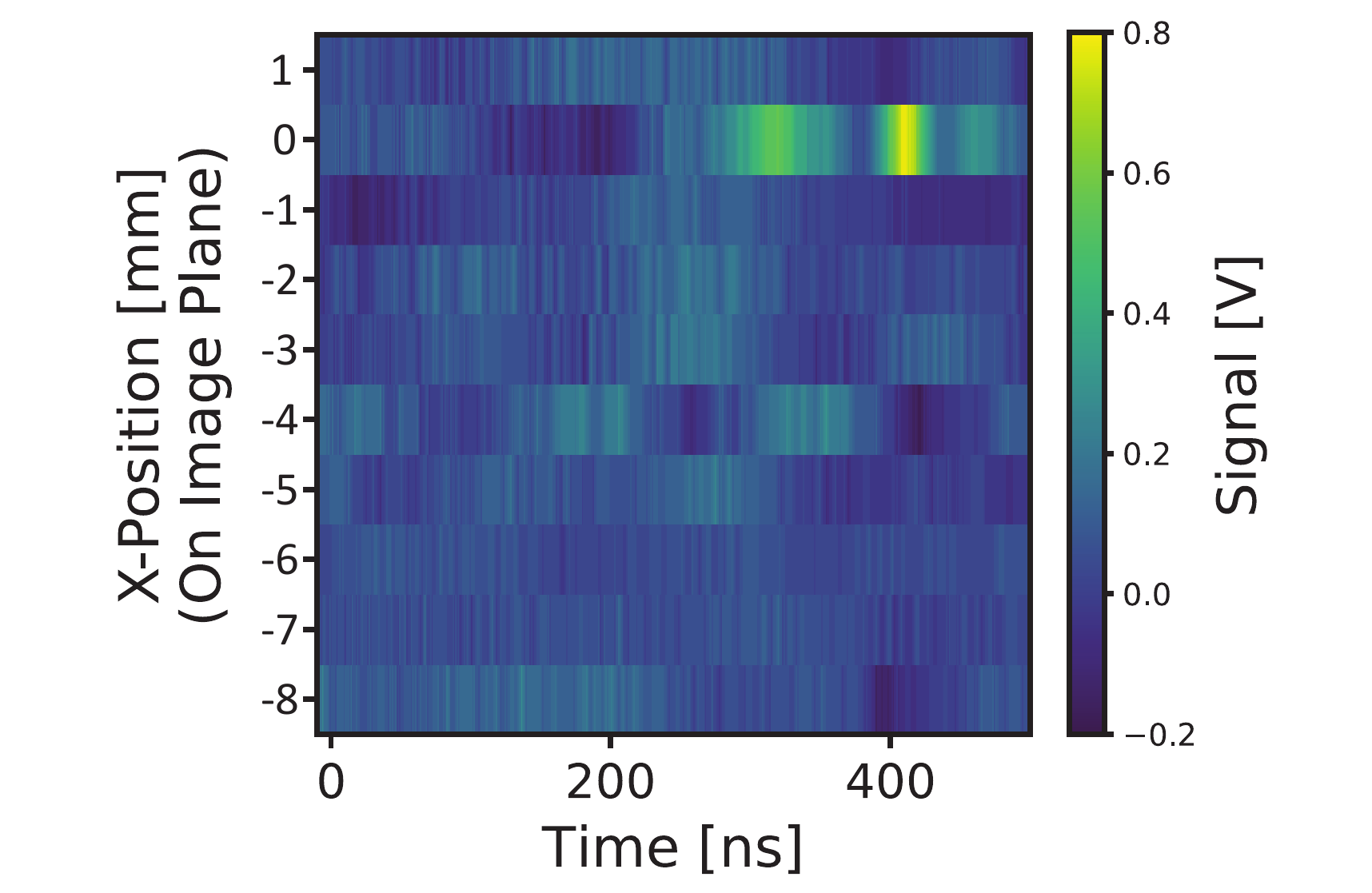}
	\caption{
		A heatmap showing the diode signals obtained in a magnetic reconnection experiment. 
	}
	\label{fig:diode_heatmap}     
\end{figure}

Figure \ref{fig:diode_heatmap} shows the diode signals obtained in the same experiment. The X-axis of the plot is time (in nanoseconds) after current start; the signal from each diode is mapped to a position on the vertical axis which corresponds to its location in the array; and the colour-map  indicates diode signal (in volts). As with the image plate data, the \SI{6}{\micro\meter} aluminium filter applied to the slit absorbed the colder emission from the inflow region so we only expected to see a signal on the diodes which imaged the reconnection layer.  

A  comparison with image plate data enabled the data from the diode array to be related to features in the object plane.  As expected, only the diode which imaged the reconnection layer (positioned at $X=\SI{0}{\milli\meter}$) shows a significant signal. This response persists during the interval 250-\SI{500}{\nano\second} after current start. The presence of multiple peaks in the data may be related to the disruption of the reconnection layer at late times. The amplitude of the peaks are  generally as expected from other diagnostics, however we currently lack the detailed knowledge of the emission spectrum from the layer which is required to make this statement more quantitative.     
 
For the purpose of this paper, the key points to take away from this section is that the emission from the layer is imaged onto the diode array and that comparing data from the array with images captured using a time integrated detector allows the signal on a particular diode to be related to regions in the experimental volume, without relying on precise alignment of the diagnostic with respect to the experimental load.

%We note that to diagnose dynamics along the layer, it have been possible to rotate the diagnostic through \SI{90}{\deg}, so the slit image showed variation along the length of the layer (in the Y-direction). Resolving structure along the length of the layer is  of interest as it could enable the of features known as plasmoids \cite{Loureiro2012, Loureiro2016}. These are islands with a closed field-topology which exhibit enhanced density and temperature, and are advected out of the layer during the course of an experiment. We have previously observed plasmoids with laser probing diagnostics in reconnection experiments on MAGPIE  \cite{Hare2017d, Hare2017b}.The combination of temporal and spatial resolution is essential for the identification of plasmoids. A time integrating diagnostic would not be useful for this purpose due to the occurrence of motion blur.      

\section{Conclusions}
We presented a design for a time-resolved X-ray imaging system to characterise X-ray self emission in HED experiments. The diagnostic incorporates two independent imaging systems, one with two-dimensional, and one with one-dimensional spatial resolution. Both imaging systems are used to record time-integrated X-ray self-emission images onto image plate. The one-dimensional imaging system was also used in combination with a linear array of 20 photodiodes to record time-resolved data. A comparison between the image-plate data from the two imaging systems enabled the features in one-dimensional images to be understood in the context of an experimental morphology which varied in two dimensions. This diagnostic is applicable to HED experiments which are either extended in one dimension or have a quasi one dimensional geometry.

\section*{Acknowledgements}
This work was supported by the U.S. Department of Energy (DOE) under Award Nos. DE-SC0020434, DE-NA0003764, DE-F03-02NA00057, and DE-SC-0001063; and by the U.S. Defence Threat Reduction Agency (DTRA) under award number HDTRA1-20-1-0001.

\section*{Data Availability}
The data that support the findings of this study are available from the corresponding author upon reasonable request.

\bibliography{Bibliography}% Produces the bibliography via BibTeX.

\end{document}